\documentclass[12pt,epsf]{article}
\usepackage{amssymb,amsmath}
\usepackage{graphicx}

\newcommand{\be}{\begin{equation}}
\newcommand{\ee}{\end{equation}}
\newcommand{\bea}{\begin{eqnarray}}
\newcommand{\eea}{\end{eqnarray}}
\newcommand{\beas}{\begin{eqnarray*}}
\newcommand{\eeas}{\end{eqnarray*}}
\newcommand{\ba}{\begin{array}}
\newcommand{\ea}{\end{array}}

\newcommand{\nbox}{{\,\lower0.9pt\vbox{\hrule \hbox{\vrule height 0.2 cm \hskip 0.19 cm \vrule height 0.2 cm}\hrule}\,}}

\def\href#1#2{#2}
\input{tcilatex}

\begin{document}

\begin{titlepage}
\hfill
\vbox{
    \halign{#\hfil         \cr
           } 
      }  

\hbox to \hsize{{}\hss \vtop{ \hbox{MCTP-08-51}

}}

\vspace*{20mm}
\begin{center}
{\Large \bf M2-M5 Systems in ${\cal N}=6$ Chern-Simons Theory\\}

\vspace*{15mm} \vspace*{1mm} Kentaro Hanaki{\footnote {e-mail:
hanaki@umich.edu}} and Hai Lin{\footnote {e-mail:
hailin@umich.edu}}

\vspace*{1cm}

{\it Department of Physics and Michigan Center for Theoretical Physics \\
University of Michigan, Ann Arbor, MI 48109, USA \\}

\vspace*{1cm}
\end{center}

\begin{abstract}
We study two aspects of M5-branes in ${\cal N}=6$ $U(N)\times
U(N)$ Chern-Simons gauge theory. We first examine multiple
M2-branes ending on a M5-brane. We study Basu-Harvey type cubic
equations, fuzzy funnel configurations, and derive the M5-brane
tension from the ${\cal N}=6$ theory. We also find a limit in
which the above M2-M5 system reduces to a D2-D4 system and we
recover the Nahm equation from the ${\cal N}=6$ theory. We then
examine domain wall configurations in mass-deformed ${\cal N}=6$
theory with a manifest $SU(2)\times SU(2)\times U(1)$ global
symmetry. We derive tensions of domain walls connecting between
arbitrary M5-brane vacua of the deformed theory and observe their
consistency with gravity dual expectations.

\end{abstract}

\end{titlepage}

\vskip 1cm

\section{Introduction}

Multiple M2-brane theory with a manifest $SO(8)$ R-symmetry was shown \cite%
{Bagger:2007jr, Gustavsson:2007vu} to be consistent with a totally
antisymmetric 3-algebraic description. The only finite dimensional Euclidean
3-algebra assuming total antisymmetry was based on the $so(4)$ 3-algebra
with a quantized 4-index structure constant \cite%
{Papadopoulos:2008sk,Bandres:2008vf}. The corresponding theory can be
presented as a $SU(2)_{k}\times SU(2)_{-k}~$Chern-Simons gauge theory \cite%
{VanRaamsdonk:2008ft, Berman:2008be} coupled to 8 scalars and 8 fermions in
bi-fundamental representations \cite{VanRaamsdonk:2008ft}. The theory was
shown to arise from two M2-branes moving in an orbifold of transverse $R^{8}$
space \cite{Distler:2008mk}, and reduce to a maximally supersymmetric
multiple D2-brane theory in a large $k$ and large scalar vev limit \cite%
{Mukhi:2008ux, Distler:2008mk}. One-loop corrections to the couplings were
considered in \cite{Gustavsson:2008bf}. Generalizations to include an
arbitrary higher rank non-abelian gauge symmetry lead to the Lorentzian
3-algebra \cite{Gomis:2008uv, DeMedeiros:2008zm}, but the corresponding
theory contains ghost degrees of freedom due to the Lorentzian signature 
\cite{Gomis:2008uv}. A ghost-removing procedure turns the theory into that
of a dual description of the 3d maximally supersymmetric Yang-Mills theory 
\cite{Bandres:2008kj, Cecotti:2008qs, Honma:2008jd, Bergshoeff:2008ix}.
Besides, infinite dimensional 3-algebras also exist \cite{Ho:2008nn,
Lin:2008qp}.

An alternative method to include a higher rank gauge symmetry was obtained
very recently by considering the $U(N)_{k}\times U(N)_{-k}~$Chern-Simons
gauge theory coupled to four $\mathcal{N}=2$ superfields in bi-fundamental
representations \cite{Aharony:2008ug}. The Lagrangian of the theory exhibits
a manifest $SU(4)$ R-symmetry \cite{Benna:2008zy,
Bagger:2008se,Bandres:2008ry}, see also \cite{Terashima:2008sy,
Gaiotto:2008cg, Hosomichi:2008jb, Gomis:2008vc}, and was proposed to arise
from multiple M2-branes moving in a $Z_{k}~$quotient of the transverse $%
R^{8} $ space \cite{Aharony:2008ug}. The theory was also shown to be
consistent with a 3-algebraic description with a less antisymmetric
structure constant \cite{Bagger:2008se}.

The present paper is motivated by trying to understand better the properties
of this new $\mathcal{N}=6~$theory. A nice feature of the previous $\mathcal{%
N}=8~$theory is that it admits the Basu-Harvey equation \cite%
{Basu:2004ed,Krishnan:2008zm} with a $SO(4)$ symmetry, and as a result,
there are fuzzy funnel configurations describing multiple M2-branes
gradually ending on a M5-brane wrapping a fuzzy 3-sphere. Another nice
feature is that the $\mathcal{N}=8~$theory also admits a mass deformation
keeping a $SO(4)\times SO(4)$~global symmetry \cite{Pope:2003jp,
Bena:2004jw, Lin:2004nb, Lin:2005nh, Gomis:2008cv}, which has multiple
M5-brane vacua charaterized by M5-branes wrapping concentric fuzzy 3-spheres
in two possible orthogonal $R^{4}~$spaces. In this paper, we will study
these two aspects in the context of the $\mathcal{N}=6~$theory.

The organization of this paper is as follows. In section 2.1, we derive
Basu-Harvey type equations by the method of forming perfect squares
combining the kinetic terms with F-terms or D-terms. Related discussion but
with slightly different methods was given in \cite%
{Gomis:2008vc,Terashima:2008sy}. In section 2.2, we analyze properties of
the fuzzy funnel solutions and derive the M5-brane tension from the $%
\mathcal{N}=6~$theory. In section 2.3, we show a limit that the above
Basu-Harvey equations reduce to Nahm equations describing D2-D4 systems,
thus giving another consistency check. In section 3.1, we derive domain wall
equations in the mass-deformed $\mathcal{N}=6~$theory keeping a $SU(2)\times
SU(2)\times U(1)$~global symmetry \cite{Gomis:2008vc}. In section 3.2, we
analyze properties of the domain walls and compute their tensions, which are
consistent with gravity dual descriptions in terms of M5-brane actions. In
section 4, we briefly draw conclusions.

\section{Basu-Harvey configurations and M2-M5 system}

\vspace{1pt}\label{basu_2}

\vspace{1pt}

\subsection{Bogomol'nyi completion}

\label{basu_2.1}

We begin by examining the bosonic potential in $\mathcal{N}=6~U(N)\times
U(N) $ Chern-Simons theory and expressing it as a sum of several perfect
squares. We basically follow the notation of \cite{Benna:2008zy}, but use a
different normalization condition for $U(N)$ generators $\mbox{tr}%
(T^{a}T^{b})=(1/2)\delta ^{ab}$. In this notation, the potential can be
rewritten as 
\begin{eqnarray}
V_{scalar} &=&V_{D}+V_{F}  \notag \\
&=&\frac{4\pi ^{2}}{k^{2}}\mbox{tr}~(|Z^{A}Z_{A}^{\dagger
}Z^{B}-Z^{B}Z_{A}^{\dagger }Z^{A}-W^{\dagger
A}W_{A}Z^{B}+Z^{B}W_{A}W^{\dagger A}|^{2}  \notag \\
&&+|W^{\dagger A}W_{A}W^{\dagger B}-W^{\dagger B}W_{A}W^{\dagger
A}-Z^{A}Z_{A}^{\dagger }W^{\dagger B}+W^{\dagger B}Z_{A}^{\dagger
}Z^{A}|^{2})  \notag \\
&&+\frac{16\pi ^{2}}{k^{2}}\mbox{tr}\left( |\epsilon _{AC}\epsilon
^{BD}W_{B}Z^{C}W_{D}|^{2}+|\epsilon ^{AC}\epsilon
_{BD}Z^{B}W_{C}Z^{D}|^{2}\right) ,
\end{eqnarray}%
where $Z^{A},W^{\dagger A},~A=1,2~$are the lowest components of four $%
\mathcal{N}=2~$superfields respectively, and are all in the $(N,\overline{N}%
)~$representations and have overall $U(1)$ charges +1. \vspace{1pt}The
classical vacuum moduli space can be determined by demanding all the squares
to be zero simultaneously. In this theory there is an additional residual $%
Z_{k}~$symmetry which orbifolds the moduli space.

Next we want to consider Basu-Harvey type BPS equations, which have the
dependence of only one of the spatial worldvolume coordinate, say $x^{2}=s$.
The equations can be obtained by combining the kinetic terms and potential
terms in the Hamiltonian and rewriting it as a sum of perfect squares plus
some topological terms.

There are two ways to make combinations. If we combine the kinetic terms
with F-term potentials, we obtain 
\begin{eqnarray}
H &=&\int dx^{1}ds~\mbox{tr}(|\partial _{s}W^{\dagger A}|^{2}+|\partial
_{s}Z^{A}|^{2}+V_{scalar})  \notag \\
&=&\int dx^{1}ds~\mbox{tr}(|\partial _{s}W^{\dagger A}-\frac{4\pi }{k}%
\epsilon ^{AC}\epsilon _{BD}Z^{B}W_{C}Z^{D}|^{2}+|\partial _{s}Z^{A}-\frac{%
4\pi }{k}\epsilon ^{AC}\epsilon _{BD}W^{\dagger B}Z_{C}^{\dagger }W^{\dagger
D}|^{2}  \notag \\
&&+\frac{4\pi ^{2}}{k^{2}}|Z^{A}Z_{A}^{\dagger }Z^{B}-Z^{B}Z_{A}^{\dagger
}Z^{A}-W^{\dagger A}W_{A}Z^{B}+Z^{B}W_{A}W^{\dagger A}|^{2}  \notag \\
&&+\frac{4\pi ^{2}}{k^{2}}|W^{\dagger A}W_{A}W^{\dagger B}-W^{\dagger
B}W_{A}W^{\dagger A}-Z^{A}Z_{A}^{\dagger }W^{\dagger B}+W^{\dagger
B}Z_{A}^{\dagger }Z^{A}|^{2})  \notag \\
&&+\frac{4\pi }{k}\epsilon _{AC}\epsilon ^{BD}\int dx^{1}~\mbox{tr}%
(Z^{A}W_{B}Z^{C}W_{D}+W^{\dagger A}Z_{B}^{\dagger }W^{\dagger
C}Z_{D}^{\dagger })  \label{Fterm_combine01}
\end{eqnarray}%
or, if the kinetic terms are combined with D-term potentials, we get:%
\begin{eqnarray}
H &=&\int dx^{1}ds~\mbox{tr}(|\partial _{s}W^{\dagger A}+\frac{2\pi }{k}%
(W^{\dagger B}W_{B}W^{\dagger A}-W^{\dagger A}W_{B}W^{\dagger
B}-Z^{B}Z_{B}^{\dagger }W^{\dagger A}+W^{\dagger A}Z_{B}^{\dagger
}Z^{B})|^{2}  \notag \\
&&+|\partial _{s}Z^{A}+\frac{2\pi }{k}(Z^{B}Z_{B}^{\dagger
}Z^{A}-Z^{A}Z_{B}^{\dagger }Z^{B}-W^{\dagger
B}W_{B}Z^{A}+Z^{A}W_{B}W^{\dagger B})|^{2}  \notag \\
&&+\frac{16\pi ^{2}}{k^{2}}|\epsilon _{AC}\epsilon ^{BD}W_{B}Z^{C}W_{D}|^{2}+%
\frac{16\pi ^{2}}{k^{2}}|\epsilon ^{AC}\epsilon _{BD}Z^{B}W_{C}Z^{D}|^{2}) 
\notag \\
&&+\frac{\pi }{k}\int dx^{1}~\mbox{tr}(W_{A}W^{\dagger A}W_{B}W^{\dagger
B}-W^{\dagger A}W_{A}W^{\dagger B}W_{B}+2W^{\dagger
A}W_{A}Z^{B}Z_{B}^{\dagger }  \notag \\
&&-2W_{A}W^{\dagger A}Z_{B}^{\dagger }Z^{B}+Z_{A}^{\dagger
}Z^{A}Z_{B}^{\dagger }Z^{B}-Z^{A}Z_{A}^{\dagger }Z^{B}Z_{B}^{\dagger }).
\label{Dterm_combine01}
\end{eqnarray}%
In each case, the last term is topological and doesn't affect the dynamics
in the bulk. So we get a set of BPS equations, which minimizes the energy in
a given topological sector: 
\begin{equation}
\partial _{s}W^{\dagger A}-\frac{4\pi }{k}\epsilon ^{AC}\epsilon
_{BD}Z^{B}W_{C}Z^{D}=0
\end{equation}%
\begin{equation}
\partial _{s}Z^{A}-\frac{4\pi }{k}\epsilon ^{AC}\epsilon _{BD}W^{\dagger
B}Z_{C}^{\dagger }W^{\dagger D}=0
\end{equation}%
\begin{equation}
Z^{A}Z_{A}^{\dagger }Z^{B}-Z^{B}Z_{A}^{\dagger }Z^{A}-W^{\dagger
A}W_{A}Z^{B}+Z^{B}W_{A}W^{\dagger A}=0
\end{equation}%
\begin{equation}
W^{\dagger A}W_{A}W^{\dagger B}-W^{\dagger B}W_{A}W^{\dagger
A}-Z^{A}Z_{A}^{\dagger }W^{\dagger B}+W^{\dagger B}Z_{A}^{\dagger }Z^{A}=0
\end{equation}%
for the F-term combination, and 
\begin{equation}
\partial _{s}W^{\dagger A}+\frac{2\pi }{k}(W^{\dagger B}W_{B}W^{\dagger
A}-W^{\dagger A}W_{B}W^{\dagger B}-Z^{B}Z_{B}^{\dagger }W^{\dagger
A}+W^{\dagger A}Z_{B}^{\dagger }Z^{B})=0
\end{equation}%
\begin{equation}
\partial _{s}Z^{A}+\frac{2\pi }{k}(Z^{B}Z_{B}^{\dagger
}Z^{A}-Z^{A}Z_{B}^{\dagger }Z^{B}-W^{\dagger
B}W_{B}Z^{A}+Z^{A}W_{B}W^{\dagger B})=0  \label{Z_basu_01}
\end{equation}%
\begin{equation}
\epsilon _{AC}\epsilon ^{BD}W_{B}Z^{C}W_{D}=\epsilon ^{AC}\epsilon
_{BD}Z^{B}W_{C}Z^{D}=0
\end{equation}%
for the D-term combination, respectively. The topological term gives the
energy of the configuration when the BPS equations are satisfied.

\subsection{Fuzzy funnel solution and M5-brane tension}

The new Basu-Harvey equation proposed in \cite{Terashima:2008sy,
Gomis:2008vc} can be obtained by setting two complex scalars to be zero, and
look at the non-trivial equations for the other two complex scalars. For
example, we can set $W^{\dagger A}=0,~$and $Z^{A}\neq 0~$in (\ref{Z_basu_01}%
). The scalar part of the Hamiltonian is given as a square term plus a
topological term: 
\begin{eqnarray}
H &=&\int dx^{1}ds~\mbox{tr}(|\partial _{s}Z^{A}+\frac{2\pi }{k}%
(Z^{B}Z_{B}^{\dagger }Z^{A}-Z^{A}Z_{B}^{\dagger }Z^{B})|^{2})  \notag \\
&&+\frac{\pi }{k}\int dx^{1}ds~\partial _{s}\mbox{tr}(Z_{A}^{\dagger
}Z^{A}Z_{B}^{\dagger }Z^{B}-Z^{A}Z_{A}^{\dagger }Z^{B}Z_{B}^{\dagger }).
\end{eqnarray}%
The first line gives a pair of BPS equations%
\begin{equation}
\partial _{s}Z^{A}+\frac{2\pi }{k}(Z^{B}Z_{B}^{\dagger
}Z^{A}-Z^{A}Z_{B}^{\dagger }Z^{B})=0,  \label{Z_basu_02}
\end{equation}%
where $A,B=1,2.~$As opposed to the original Basu-Harvey equation in \cite%
{Basu:2004ed} which has a manifest $SO(4)$ symmetry, the equation (\ref%
{Z_basu_02}) has a manifest $SU(2)\times U(1)~$symmetry. As was argued in 
\cite{Terashima:2008sy}, this equation preserves half of the supersymmetries
of the theory. For a configuration on which this equation is satisfied, the
energy of the system is given by%
\begin{eqnarray}
E &=&\frac{\pi }{k}\int dx^{1}~\mbox{tr}(Z_{A}^{\dagger }Z^{A}Z_{B}^{\dagger
}Z^{B}-Z^{A}Z_{A}^{\dagger }Z^{B}Z_{B}^{\dagger }) \\
&=&2\int dsdx^{1}\mbox{tr}(\partial _{s}Z_{A}^{\dagger }\partial _{s}Z^{A}).
\label{Basu_energy01}
\end{eqnarray}%
We used the BPS equation (\ref{Z_basu_02}) to obtain the second line.

To solve the BPS equation (\ref{Z_basu_02}), we may separate the $s$%
-dependent and independent part: 
\begin{equation}
Z^{A}=f(s)G^{A},\quad f(s)=\sqrt{\frac{k}{4\pi s}},  \label{funnel_profile01}
\end{equation}%
where $G^{A}$s are $N\times \overline{N}~$matrices satisfying 
\begin{equation}
G^{A}=G^{B}G_{B}^{\dagger }G^{A}-G^{A}G_{B}^{\dagger }G^{B}.
\end{equation}%
This equation is solved in \cite{Gomis:2008vc} (see also \cite%
{Terashima:2008sy}). One can diagonalize $G_{1}^{\dagger }~$using the $%
U(N)\times U(N)~$transformations and find that the other matrix $%
G_{2}^{\dagger }~$must be off-diagonal. The $G_{A}^{\dagger }$s~have some
nice properties: For a $N$ dimensional irreducible solution, 
\begin{eqnarray}
(G_{1}^{\dagger })_{m,n} &=&\sqrt{m-1}\delta _{m,n},~~~(G_{2}^{\dagger
})_{m,n}=\sqrt{N-m}\delta _{m+1,n},  \label{G_A_01} \\
G^{1}G_{1}^{\dagger } &=&\mathrm{diag~}(0,1,2,\ ...\ ,N-1)=~G_{1}^{\dagger
}G^{1} \\
G^{2}G_{2}^{\dagger } &=&\mathrm{diag~}(N-1,N-2,~...~,1,0) \\
G_{2}^{\dagger }G^{2} &=&\mathrm{diag~}(0,N-1,N-2,~...~,1) \\
G^{A}G_{A}^{\dagger } &=&(N-1)\mathbf{1}_{N\times N},~~~~\mathrm{tr}%
(G^{A}G_{A}^{\dagger })=N(N-1).  \label{G_A_trace01}
\end{eqnarray}%
The eigenvalues of the matrices $G^{1}G_{1}^{\dagger }~$and $%
G^{2}G_{2}^{\dagger }~$may be interpreted as the squares of the radial
positions of the points on a fuzzy 3-sphere projected onto 2 complex planes,
respectively. Since there is a overall $Z_{k}~$residual symmetry, the
solution would describe a fuzzy $S^{3}/Z_{k}.$

The energy formula (\ref{Basu_energy01}) is expressed in terms of fields $%
Z^{A}$, which is of mass dimension $1/2$ and does not have the correct mass
dimension $-1$ as a spatial coordinate. The correct normalization should
reproduce the scalar kinetic term of the form, 
\begin{equation}
S_{kinetic}=-T_{2}\int d^3x \mbox{tr}(\partial _{\mu }X_{A}^{\dagger
}\partial ^{\mu }X^{A}),  \label{normal}
\end{equation}%
where $T_{2}$ is the M2-brane tension and $X^{A}$ is the (complexified)
spatial coordinate. This implies that we should relate $X^{A}$ and $Z^{A}$
by 
\begin{equation}
X^{A}=\sqrt{\frac{1}{T_{2}}}Z^{A}.
\end{equation}%
Using this, we can define the radius averaged over each M2-brane as 
\begin{eqnarray}
R^{2} &=&\frac{2\mbox{tr}(X_{A}^{\dagger }X^{A})}{N}=\frac{2(N-1)}{T_{2}}%
f^{2} \\
&=&\frac{k(N-1)}{2\pi T_{2}}~\cdot \frac{1}{s}
\end{eqnarray}%
The factor of two in the numerator comes from our normalization condition $%
\mbox{tr}(T^{a}T^{b})=(1/2)\delta ^{ab}$. The radius vanishes for $N=1$, and
there are non-trivial fuzzy 3-spheres only for $N \geq 2$.

Combining all the above results, after some algebra, we obtain 
\begin{eqnarray}
E &=&\frac{T_{2}^{2}}{2\pi }\frac{N}{N-1}\int dx^{1}\left( \frac{2\pi ^{2}}{k%
}\right) R^{3}dR \\
&=&\frac{T_{2}^{2}}{2\pi }\frac{N}{N-1}\int d^{5}x.
\end{eqnarray}%
The factor $k$ in the denominator represents the fact that this M5-brane is
divided by the $Z_{k}$ orbifold action, and $\frac{2\pi ^{2}}{k}~$is the
volume of an $S^{3}/Z_{k}$ with a unit radius. So the M5-brane wraps an $%
S^{3}/Z_{k}.~$The M5-brane tension predicted from the $\mathcal{N}=6~$theory
is 
\begin{equation}
T_{5}=\frac{T_{2}^{2}}{2\pi }\frac{N}{N-1}.  \label{N=6_tension_01}
\end{equation}

The relation between M2-brane and M5-brane tension can also be derived in
different ways, by matching the M-theory and type II string theory BPS
spectrum \cite{Schwarz:1995jq}, or by applying flux and Dirac quantization
rules in eleven dimensions \cite{Duff:1995wd}: 
\begin{equation}
T_{5}=\frac{T_{2}^{2}}{2\pi }.  \label{M5tension}
\end{equation}%
We see that for large $N$ including the numerical coefficient, (\ref%
{N=6_tension_01})\ exactly agrees with the known result (\ref{M5tension}).
The $1/N$ deviation is due to the fuzziness of the 3-sphere in the finite $N$
regime, and will disappear in the continuum limit for the fuzzy 3-sphere.

\subsection{Basu-Harvey equations and reduction to Nahm equations}

\label{basu_2.2} 

In this section we take a limit in which M2-brane theory reduces to D2-brane
theory \cite{Pang:2008hw, Li:2008ya} and show that the Basu-Harvey equation (%
\ref{Z_basu_02}) studied in the last section reduces to a Nahm equation,
which describes D2-D4 system.

We take a diagonal expectation value in one of the direction, for example,
the direction labelled by 3 and expand the fields around the vacuum: 
\begin{eqnarray}
Z^{1} &=&(x^{10}+ix^{20})T^{0}+X^{1}+iX^{2}\quad  \label{large_v_X} \\
Z^{2} &=&((v+x^{30})+ix^{40})T^{0}+X^{3}+iX^{4}
\end{eqnarray}%
Here, $x$'s represent the $U(1)$ part and $T^{0}=\frac{1}{\sqrt{2N}}\mathbf{%
1~}$for normalization purpose, $\mbox{tr}(T^{0}T^{0})=1/2$. $X$'s take value
on $SU(N)$.$\mathbf{~}$We take $N$ and $v/k$ finite and fixed, and suppose $v
$ is large, and then we will neglect $o(1/v)~$terms in the calculation below.

By plugging (\ref{large_v_X}) into the BPS equation (\ref{Z_basu_02}), we
see that 
\begin{eqnarray}
\partial _{s}Z^{2} &=&\frac{2\pi }{k}(Z^{2}Z_{1}^{\dagger
}Z^{1}-Z^{1}Z_{1}^{\dagger }Z^{2}) \\
&=&\frac{2\pi v}{k\sqrt{2N}}[Z_{1}^{\dagger },Z^{1}] \\
&=&\frac{4\pi v}{k\sqrt{2N}}i[X^{1},X^{2}]
\end{eqnarray}%
$U(1)$ part decouples from the equations and we simply set them to zero. $%
SU(N)$ part implies 
\begin{equation}
\partial _{s}X^{3}=\frac{4\pi v}{k\sqrt{2N}}i[X^{1},X^{2}],\quad \partial
_{s}X^{4}=0
\end{equation}%
where we compared hermitian and anti-hermitian parts respectively.

In the same way, we can calculate the other component equation 
\begin{eqnarray}
\partial _{s}Z^{1} &=&\frac{2\pi }{k}(Z^{1}Z_{2}^{\dagger
}Z^{2}-Z^{2}Z_{2}^{\dagger }Z^{1}) \\
&=&\frac{2\pi v}{k\sqrt{2N}}[Z^{1},Z_{2}^{\dagger }+Z^{2}] \\
&=&\frac{4\pi v}{k\sqrt{2N}}([X^{1},X^{3}]+i[X^{2},X^{3}])
\end{eqnarray}%
So we get%
\begin{eqnarray}
\partial _{s}X^{1} &=&\frac{4\pi v}{k\sqrt{2N}}i[X^{2},X^{3}] \\
\partial _{s}X^{2} &=&\frac{4\pi v}{k\sqrt{2N}}i[X^{3},X^{1}]
\end{eqnarray}

Combining the above results, we get 
\begin{equation}
\partial _{s}X^{i}=i\frac{1}{2}g_{YM}\epsilon ^{ijk}[X^{j},X^{k}]
\end{equation}%
where $i,j,k=1,2,3$ and $\epsilon ^{ijk}$ is the totally antisymmetric
tensor. By using $g_{YM}=4\pi v/k\sqrt{2N}$ as in the M2 to D2 reduction 
\cite{Pang:2008hw, Li:2008ya} for the $\mathcal{N}=6$ theory, we get the
Nahm equation with the exact coefficient, in the large $v$ and large $k$
limit, with $N~$and $v/k~$fixed and finite.~This describes$~$multiple
D2-branes ending on a D4-brane wrapping an $S^{2}$, and the reduction
process makes an $S^{3}/Z_{k}~$reducing to an $S^{2}~$that the D4-brane
wraps.

\section{Domain wall configurations and M2-M5 system}

\label{domain}

\subsection{Domain wall equations}

\vspace{1pt}\label{domain_equation}\vspace{1pt}

In this section, we turn to the discussion of another aspect of the
M5-branes in the $\mathcal{N}=6$ theory. \vspace{1pt}For the $\mathcal{N}=8~$%
M2-brane theory on flat space, we can turn on four fermion mass terms, which
preserve at least $\mathcal{N}=2~$supersymmetry. The most symmetric mass
deformation is the one preserving a $SO(4)\times SO(4)~$symmetry \cite%
{Pope:2003jp, Bena:2004jw, Lin:2004nb} and a $SU(2|2)\times SU(2|2)$
superalgebra. In this case, M5-branes can wrap either of the two geometric $%
S^{3}$s in orthogonal $R^{4}$s.$~$

In the case of $\mathcal{N}=6$ formulation, the most symmetric mass
deformation turns out to preserve a manifest $SU(2)\times SU(2)\times U(1)~$%
symmetry \cite{Gomis:2008vc} (see also related discussion \cite%
{Hosomichi:2008jb,Hosomichi:2008jd}) and we expect to have a $SU(2|2)\times
SU(1|1)~$superalgebra. While, in this case, M5-branes can wrap either of two
possible geometric ($S^{3}/Z_{k})$s, where the $Z_{k}$ action is due to the
residual symmetry, which squash the 3-spheres along their Hopf fiber
directions while maintaining a manifest $SU(2)\times U(1)~$symmetry, as in (%
\ref{Z_basu_02}).

We can turn on a D-term deformation corresponding to adding a FI term as
found in \cite{Gomis:2008vc}. In our notation, we have the deformed potential%
\begin{eqnarray}
V_{scalar} &=&V_{D}+V_{F}  \notag \\
&=&\frac{4\pi ^{2}}{k^{2}}\mbox{tr}(|-\frac{k}{2\pi }\mu
Z^{B}+Z^{A}Z_{A}^{\dagger }Z^{B}-Z^{B}Z_{A}^{\dagger }Z^{A}-W^{\dagger
A}W_{A}Z^{B}+Z^{B}W_{A}W^{\dagger A}|^{2}  \notag \\
&&+|-\frac{k}{2\pi }\mu W^{\dagger B}+W^{\dagger A}W_{A}W^{\dagger
B}-W^{\dagger B}W_{A}W^{\dagger A}-Z^{A}Z_{A}^{\dagger }W^{\dagger
B}+W^{\dagger B}Z_{A}^{\dagger }Z^{A}|^{2})  \notag \\
&&+\frac{16\pi ^{2}}{k^{2}}\mbox{tr}\left( |\epsilon _{AC}\epsilon
^{BD}W_{B}Z^{C}W_{D}|^{2}+|\epsilon ^{AC}\epsilon
_{BD}Z^{B}W_{C}Z^{D}|^{2}\right)
\end{eqnarray}%
where $\mu $ is a canonical mass parameter.

We perform the Bogomol'nyi completion combining the kinetic terms and
D-terms similar to (\ref{Dterm_combine01}), and we get%
\begin{eqnarray}
H &=&\int dx^{1}ds\mbox{tr}(|\partial _{s}W^{\dagger A}-\mu W^{\dagger A}+%
\frac{2\pi }{k}(W^{\dagger B}W_{B}W^{\dagger A}-W^{\dagger A}W_{B}W^{\dagger
B}  \notag \\
&&-Z^{B}Z_{B}^{\dagger }W^{\dagger A}+W^{\dagger A}Z_{B}^{\dagger
}Z^{B})|^{2}  \notag \\
&&+|\partial _{s}Z^{A}-\mu Z^{A}+\frac{2\pi }{k}(Z^{B}Z_{B}^{\dagger
}Z^{A}-Z^{A}Z_{B}^{\dagger }Z^{B}-W^{\dagger
B}W_{B}Z^{A}+Z^{A}W_{B}W^{\dagger B})|^{2}  \notag \\
&&+\frac{16\pi ^{2}}{k^{2}}|\epsilon _{AC}\epsilon ^{BD}W_{B}Z^{C}W_{D}|^{2}+%
\frac{16\pi ^{2}}{k^{2}}|\epsilon ^{AC}\epsilon _{BD}Z^{B}W_{C}Z^{D}|^{2}) 
\notag \\
&&+\frac{\pi }{k}\int dx^{1}~\mbox{tr}(W_{A}W^{\dagger A}W_{B}W^{\dagger
B}-W^{\dagger A}W_{A}W^{\dagger B}W_{B}+2W^{\dagger
A}W_{A}Z^{B}Z_{B}^{\dagger }  \notag \\
&&-2W_{A}W^{\dagger A}Z_{B}^{\dagger }Z^{B}+Z_{A}^{\dagger
}Z^{A}Z_{B}^{\dagger }Z^{B}-Z^{A}Z_{A}^{\dagger }Z^{B}Z_{B}^{\dagger }) 
\notag \\
&&+\int dx^{1}~\mathrm{tr}(\mu W^{\dagger A}W_{A}+\mu Z^{A}Z_{A}^{\dagger })
\end{eqnarray}%
New boundary topological terms are produced at the same time when the BPS
equations are modified.

The BPS domain wall equations are 
\begin{equation}
\partial _{s}W^{\dagger A}-\mu W^{\dagger A}+\frac{2\pi }{k}(W^{\dagger
B}W_{B}W^{\dagger A}-W^{\dagger A}W_{B}W^{\dagger B}-Z^{B}Z_{B}^{\dagger
}W^{\dagger A}+W^{\dagger A}Z_{B}^{\dagger }Z^{B})=0  \label{domain_eqn_06}
\end{equation}%
\begin{equation}
\partial _{s}Z^{A}-\mu Z^{A}+\frac{2\pi }{k}(Z^{B}Z_{B}^{\dagger
}Z^{A}-Z^{A}Z_{B}^{\dagger }Z^{B}-W^{\dagger
B}W_{B}Z^{A}+Z^{A}W_{B}W^{\dagger B})=0  \label{domain_eqn_07}
\end{equation}%
\begin{equation}
\epsilon _{AC}\epsilon ^{BD}W_{B}Z^{C}W_{D}=\epsilon ^{AC}\epsilon
_{BD}Z^{B}W_{C}Z^{D}=0 .  \label{domain_eqn_08}
\end{equation}%
The equations are modified by just adding the linear terms.

\subsection{Domain wall solutions and their tensions}

\label{domain_tension}\vspace{1pt}

In this section we discuss solutions of these domain wall configurations and
derive their tensions. Setting $W^{\dagger A}=0~$in equations (\ref%
{domain_eqn_06})-(\ref{domain_eqn_08}), we need to solve 
\begin{equation}
\partial _{s}Z^{A}-\mu Z^{A}+\frac{2\pi }{k}(Z^{B}Z_{B}^{\dagger
}Z^{A}-Z^{A}Z_{B}^{\dagger }Z^{B})=0  \label{domain_01}
\end{equation}

We assume the ansatz%
\begin{eqnarray}
&&Z^{A}=~h(s)G^{A},~~~~G^{A}=G^{B}G_{B}^{\dagger }G^{A}-G^{A}G_{B}^{\dagger
}G^{B} \\
&&\partial _{s}h-\mu h+\frac{2\pi }{k}h^{3}=0  \label{h_eqn_domain04}
\end{eqnarray}%
We then obtain two solutions 
\begin{eqnarray}
h_{1}(s) &=&\sqrt{\frac{k\mu }{2\pi \left( 1-e^{-2\mu s}\right) }}~ \\
h_{2}(s) &=&~\sqrt{\frac{k\mu }{2\pi \left( 1+e^{-2\mu s}\right) }}
\end{eqnarray}

The first solution $h_{1}~$describes a fuzzy funnel where $s\in (0,\infty )$%
,~and in the $\mu \rightarrow 0$ limit reproduces (\ref{funnel_profile01}).
The second solution $h_{2}~$is a domain wall solution where\ $s\in (-\infty
,\infty )$. We have 
\begin{equation}
h_{2}(-\infty )=0,~\ \ \ h_{2}(+\infty )=\sqrt{\frac{k\mu }{2\pi }}~
\end{equation}%
so this domain wall solution%
\begin{equation}
Z^{A}=~~\sqrt{\frac{k\mu }{2\pi \left( 1+e^{-2\mu s}\right) }}G^{A}
\end{equation}%
connects a trivial vacuum with a nontrivial fuzzy sphere vacuum $\sqrt{\frac{%
k\mu }{2\pi }}G^{A}$.

The non-vanishing boundary terms when $W^{\dagger A}=0$ are 
\begin{eqnarray}
H &=&\int dx^{1}ds\partial _{s}\mathrm{tr}(\mu Z^{A}Z_{A}^{\dagger })+\frac{%
\pi }{k}\int dx^{1}ds\partial _{s}\mbox{tr}(Z_{A}^{\dagger
}Z^{A}Z_{B}^{\dagger }Z^{B}-Z^{A}Z_{A}^{\dagger }Z^{B}Z_{B}^{\dagger }) \\
&=&\int dx^{1}~\mathrm{tr}(\frac{1}{2}\mu Z^{A}Z_{A}^{\dagger })|_{s=-\infty
}^{s=\infty }=2\int dx^{1}ds~\mathrm{tr}(\partial _{s}Z^{A}\partial
_{s}Z_{A}^{\dagger })  \label{boundary terms_after eqn} \\
&=&\int dx^{1}(\frac{k\mu ^{2}}{4\pi })\mbox{tr}(G^{A}G_{A}^{\dagger
})|_{s=-\infty }^{s=\infty }  \label{general_domain 04} \\
&=&\int dx^{1}\frac{k}{4\pi }\mu ^{2}N(N-1)  \label{general domain 05}
\end{eqnarray}%
where in deriving the second line in (\ref{boundary terms_after eqn}) we
have used the equation of motion (\ref{domain_01}) to simplify%
\begin{equation}
\frac{\pi }{k}(Z^{A}Z_{B}^{\dagger }Z^{B}Z_{A}^{\dagger
}-Z^{B}Z_{B}^{\dagger }Z^{A}Z_{A}^{\dagger })=-\frac{1}{2}\mu
Z^{A}Z_{A}^{\dagger }+\frac{1}{2}(\partial _{s}Z^{A})Z_{A}^{\dagger }
\label{eqn_domain02}
\end{equation}%
and used the fact that $\frac{1}{2}(\partial _{s}Z^{A})Z_{A}^{\dagger }$
vanishes for both $s=-\infty ~$and $s=\infty .$

Thereby the tension of this domain wall is 
\begin{equation}
\tau =\frac{k}{4\pi }\mu ^{2}N(N-1)  \label{simplest_domain02}
\end{equation}%
It agrees with other results for slightly different theories as discussed in 
\cite{Bagger:2007jr}, and the second ref. in \cite{Gomis:2008cv}.

Since (\ref{eqn_domain02}),(\ref{boundary terms_after eqn}) are the general
results for general domain wall solutions, we see that the expression (\ref%
{general_domain 04}) should be a general result for the tension of a domain
wall between two arbitrary vacua labelled by integers $\{N_{i}^{\prime
}|_{s=-\infty },i=1,...,p^{\prime }\},~\{N_{i}|_{s=\infty },i=1,...,p\},$ in
which the integers label the dimensions of irreducible solutions of the $%
p^{\prime }~$and $p~$diagonal-block matrices in $G^{A}|_{s=-\infty }~$and$%
~G^{A}|_{s=\infty }~$respectively.$~$The tension of the domain wall between
these two arbitrary vacua is therefore%
\begin{equation}
\tau =\frac{k\mu ^{2}}{4\pi }\sum\limits_{i=1}^{p}N_{i}(N_{i}-1)|_{s=\infty
}-\frac{k\mu ^{2}}{4\pi }\sum\limits_{i=1}^{p^{\prime }}N_{i}^{\prime
}(N_{i}^{\prime }-1)|_{s=-\infty }
\end{equation}

The dependence of (\ref{simplest_domain02}) on mass and $N$ also agrees with
the gravity dual analysis in \cite{Bena:2000zb} based on computing the
action of a M5-brane filling a 4-ball bounded by the 3-sphere on which the
M5-brane constructed from M2-branes wraps. The probe M5-brane is also along
the $R^{1,1}~$part of the M2-brane worldvolume directions. This computation
can also be performed by calculating the action of a M5-brane wrapping a $%
S^{3}$ as well as the $x_{2}~$line-segment across the fermion band at $y=0$
in the gravity geometry in \cite{Lin:2004nb},\cite{Lin:2005nh}. In this
gravity picture, it is suggestive that if the fermion band is narrow, the
M5-brane action is expected to be small.

\vspace{1pt}\vspace{1pt}

\section{Conclusions and discussion}

\label{discussion}

In this paper we have studied two problems of M5-branes in the $\mathcal{N}%
=6 $ theory. We analyzed the Basu-Harvey type equations and found evidence
that the equations describe multiple M2-branes ending on a M5-brane, which
wraps on a fuzzy 3-sphere. We derived the tension of M5-brane and it exactly
agrees with the known result in large $N$ limit. We also found that the
3-sphere is orbifolded by a $Z_{k}~$action as the volume of the M5-brane is
suppressed by $1/k.$ This is also consistent with the $SU(2)\times U(1)~$%
symmetry of the equations. We also derived the Nahm equation describing
D2-branes ending on a D4-brane wrapping an $S^{2}~$starting from the above
Basu-Harvey type equations and taking a large $k$ limit, providing further
evidence for consistency.

We then turned to another situation where M5-branes wrapping on fuzzy
3-sphere emerge as the vacua of the mass-deformed $\mathcal{N}=6$ theory. We
find domain wall solutions and computed their tensions, in agreement with
known gravity analysis, thereby adding another evidence for the existence of
the M5-branes in the $\mathcal{N}=6$ theory.

\vspace{1pt}

\vspace{0.2cm}

\section*{Acknowledgments}

It is a pleasure to thank A. Gustavsson, T. Klose, J. T. Liu, J. Maldacena,
S. Sasaki and X. Yin for helpful conversations or correspondences. This work
is supported by U.S. Department of Energy under grant DE-FG02-95ER40899, and
Michigan Center for Theoretical Physics.


\begin{thebibliography}{99}
\bibitem{Bagger:2007jr} J.~Bagger and N.~Lambert, ``Gauge Symmetry and
Supersymmetry of Multiple M2-Branes,'' Phys.\ Rev.\ D \textbf{77}, 065008
(2008) [arXiv:0711.0955 [hep-th]]; 

J.~Bagger and N.~Lambert, ``Modeling multiple M2's,'' Phys.\ Rev.\ D \textbf{%
75}, 045020 (2007)[arXiv:hep-th/0611108]; 

J.~Bagger and N.~Lambert, ``Comments On Multiple M2-branes,'' JHEP \textbf{%
0802}, 105 (2008) [arXiv:0712.3738 [hep-th]]. 

\bibitem{Gustavsson:2007vu} A.~Gustavsson, ``Algebraic structures on
parallel M2-branes,'' arXiv:0709.1260 [hep-th]; 

A.~Gustavsson, ``Selfdual strings and loop space Nahm equations,'' JHEP 
\textbf{0804}, 083 (2008) [arXiv:0802.3456 [hep-th]]. 

\bibitem{Schwarz:2004yj} J.~H.~Schwarz, ``Superconformal Chern-Simons
theories,'' JHEP \textbf{0411}, 078 (2004) [arXiv:hep-th/0411077]. 

\bibitem{Basu:2004ed} A.~Basu and J.~A.~Harvey, ``The M2-M5 brane system and
a generalized Nahm's equation,'' Nucl.\ Phys.\ B \textbf{713}, 136 (2005)
[arXiv:hep-th/0412310]. 

\bibitem{Bandres:2008vf} M.~A.~Bandres, A.~E.~Lipstein and J.~H.~Schwarz,
``N = 8 Superconformal Chern--Simons Theories,'' arXiv:0803.3242 [hep-th]. 

\bibitem{Mukhi:2008ux} S.~Mukhi and C.~Papageorgakis, ``M2 to D2,''
arXiv:0803.3218 [hep-th]. 

\bibitem{VanRaamsdonk:2008ft} M.~Van Raamsdonk, ``Comments on the
Bagger-Lambert theory and multiple M2-branes,'' arXiv:0803.3803 [hep-th]. 

\bibitem{Berman:2008be} D.~S.~Berman, L.~C.~Tadrowski and D.~C.~Thompson,
``Aspects of Multiple Membranes,'' arXiv:0803.3611 [hep-th]. 

\bibitem{Distler:2008mk} N.~Lambert and D.~Tong, ``Membranes on an
Orbifold,'' arXiv:0804.1114 [hep-th]; 

J.~Distler, S.~Mukhi, C.~Papageorgakis and M.~Van Raamsdonk, ``M2-branes on
M-folds,'' arXiv:0804.1256 [hep-th]. 

\bibitem{Gomis:2008uv} J.~Gomis, G.~Milanesi and J.~G.~Russo,
\textquotedblleft Bagger-Lambert Theory for General Lie
Algebras,\textquotedblright\ arXiv:0805.1012 [hep-th]; 

S.~Benvenuti, D.~Rodriguez-Gomez, E.~Tonni and H.~Verlinde, ``N=8
superconformal gauge theories and M2 branes,'' arXiv:0805.1087 [hep-th]; 

P.~M.~Ho, Y.~Imamura and Y.~Matsuo, ``M2 to D2 revisited,'' arXiv:0805.1202
[hep-th]. 

\bibitem{Papadopoulos:2008sk} G.~Papadopoulos, ``M2-branes, 3-Lie Algebras
and Plucker relations,'' arXiv:0804.2662 [hep-th]; 

J.~P.~Gauntlett and J.~B.~Gutowski, \textquotedblleft Constraining Maximally
Supersymmetric Membrane Actions,\textquotedblright\ arXiv:0804.3078
[hep-th]; 

P.~M.~Ho, R.~C.~Hou and Y.~Matsuo, ``3-Algebra and Multiple M2-branes,''
arXiv:0804.2110 [hep-th]; 

A.~Morozov, ``On the Problem of Multiple M2 Branes,'' arXiv:0804.0913
[hep-th]; 

U.~Gran, B.~E.~W.~Nilsson and C.~Petersson, \textquotedblleft On relating
multiple M2 and D2-branes,\textquotedblright\ arXiv:0804.1784 [hep-th]; 

E.~A.~Bergshoeff, M.~de Roo and O.~Hohm, ``Multiple M2-branes and the
Embedding Tensor,'' Class.\ Quant.\ Grav.\ \textbf{25}, 142001 (2008)
[arXiv:0804.2201 [hep-th]]. 


\bibitem{Aharony:2008ug} O.~Aharony, O.~Bergman, D.~L.~Jafferis and
J.~Maldacena, ``N=6 superconformal Chern-Simons-matter theories, M2-branes
and their gravity duals,'' arXiv:0806.1218 [hep-th]. 


\bibitem{Benna:2008zy} M.~Benna, I.~Klebanov, T.~Klose and M.~Smedback,
``Superconformal Chern-Simons Theories and $AdS_4/CFT_3$ Correspondence,''
arXiv:0806.1519 [hep-th]. 


\bibitem{Bagger:2008se} J.~Bagger and N.~Lambert, ``Three-Algebras and N=6
Chern-Simons Gauge Theories,'' arXiv:0807.0163 [hep-th]. 


\bibitem{Bandres:2008ry} M.~A.~Bandres, A.~E.~Lipstein and J.~H.~Schwarz,
``Studies of the ABJM Theory in a Formulation with Manifest SU(4)
R-Symmetry,'' arXiv:0807.0880 [hep-th]. 


\bibitem{Bandres:2008kj} M.~A.~Bandres, A.~E.~Lipstein and J.~H.~Schwarz,
``Ghost-Free Superconformal Action for Multiple M2-Branes,'' arXiv:0806.0054
[hep-th]; 

J.~Gomis, D.~Rodriguez-Gomez, M.~Van Raamsdonk and H.~Verlinde,
``Supersymmetric Yang-Mills Theory From Lorentzian Three-Algebras,''
arXiv:0806.0738 [hep-th]; 

B.~Ezhuthachan, S.~Mukhi and C.~Papageorgakis, ``D2 to D2,'' JHEP \textbf{%
0807}, 041 (2008) [arXiv:0806.1639 [hep-th]]. 


\bibitem{Cecotti:2008qs} S.~Cecotti and A.~Sen, ``Coulomb Branch of the
Lorentzian Three Algebra Theory,'' arXiv:0806.1990 [hep-th]. 


\bibitem{Honma:2008jd} Y.~Honma, S.~Iso, Y.~Sumitomo and S.~Zhang, ``Scaling
limit of N=6 superconformal Chern-Simons theories and Lorentzian
Bagger-Lambert theories,'' arXiv:0806.3498 [hep-th]. 


\bibitem{Bergshoeff:2008ix}  E.~A.~Bergshoeff, M.~de Roo, O.~Hohm and
D.~Roest,  ``Multiple Membranes from Gauged Supergravity,''  arXiv:0806.2584
[hep-th].  


\bibitem{DeMedeiros:2008zm} P.~De Medeiros, J.~Figueroa-O'Farril and
E.~Mendez-Escobar, ``Lorentzian Lie 3-algebras and their Bagger-Lambert
moduli space,'' arXiv:0805.4363 [hep-th]; 

P.~de Medeiros, J.~Figueroa-O'Farrill and E.~Mendez-Escobar, ``Metric Lie
3-algebras in Bagger-Lambert theory,'' arXiv:0806.3242 [hep-th]. 


\bibitem{Lin:2008qp} H.~Lin, ``Kac-Moody Extensions of 3-Algebras and
M2-branes,'' arXiv:0805.4003 [hep-th]; 


T.~L.~Curtright, D.~B.~Fairlie and C.~K.~Zachos, ``Ternary Virasoro - Witt
Algebra,'' arXiv:0806.3515 [hep-th]; 

T.~A.~Larsson, ``Virasoro 3-algebra from scalar densities,'' arXiv:0806.4039
[hep-th]; 

C.~Sochichiu, \textquotedblleft On Nambu-Lie 3-algebra
representations,\textquotedblright\ arXiv:0806.3520 [hep-th];%

S.~Chakrabortty, A.~Kumar and S.~Jain, ``$w_{\infty}$ 3-algebra,''
arXiv:0807.0284 [hep-th]. 

\bibitem{Gomis:2008cv} J.~Gomis, A.~J.~Salim and F.~Passerini, ``Matrix
Theory of Type IIB Plane Wave from Membranes,'' arXiv:0804.2186 [hep-th]; 

K.~Hosomichi, K.~M.~Lee and S.~Lee, ``Mass-Deformed Bagger-Lambert Theory
and its BPS Objects,'' arXiv:0804.2519 [hep-th]. 

\bibitem{Honma:2008un} Y.~Honma, S.~Iso, Y.~Sumitomo and S.~Zhang, ``Janus
field theories from multiple M2 branes,'' arXiv:0805.1895 [hep-th]. 


\bibitem{Gomis:2008vc} J.~Gomis, D.~Rodriguez-Gomez, M.~Van Raamsdonk and
H.~Verlinde, ``A Massive Study of M2-brane Proposals,'' arXiv:0807.1074
[hep-th]. 


\bibitem{Terashima:2008sy} S.~Terashima, ``On M5-branes in N=6 Membrane
Action,'' arXiv:0807.0197 [hep-th]. 


\bibitem{Gaiotto:2008cg} D.~Gaiotto, S.~Giombi and X.~Yin, ``Spin Chains in
N=6 Superconformal Chern-Simons-Matter Theory,'' arXiv:0806.4589 [hep-th]. 

\bibitem{Nishioka:2008gz} T.~Nishioka and T.~Takayanagi, ``On Type IIA
Penrose Limit and N=6 Chern-Simons Theories,'' arXiv:0806.3391 [hep-th].

\bibitem{Schwarz:1995jq} J.~H.~Schwarz, ``The power of M theory,'' Phys.\
Lett.\ B \textbf{367}, 97 (1996) [arXiv:hep-th/9510086]. 


\bibitem{Duff:1995wd} M.~J.~Duff, J.~T.~Liu and R.~Minasian,
``Eleven-dimensional origin of string / string duality: A one-loop test,''
Nucl.\ Phys.\ B \textbf{452}, 261 (1995) [arXiv:hep-th/9506126]. 


\bibitem{Pang:2008hw} Y.~Pang and T.~Wang, ``From N M2's to N D2's,''
arXiv:0807.1444 [hep-th]. 


\bibitem{Li:2008ya} T.~Li, Y.~Liu and D.~Xie, ``Multiple D2-Brane Action
from M2-Branes,'' arXiv:0807.1183 [hep-th]. 


\bibitem{Bena:2000zb} I.~Bena, ``The M-theory dual of a 3 dimensional theory
with reduced supersymmetry,'' Phys.\ Rev.\ D \textbf{62}, 126006 (2000)
[arXiv:hep-th/0004142]. 


\bibitem{Ahn:2008ya} C.~Ahn, ``Holographic Supergravity Dual to Three
Dimensional N=2 Gauge Theory,'' arXiv:0806.1420 [hep-th]. 


\bibitem{Pope:2003jp} C.~N.~Pope and N.~P.~Warner, ``A dielectric flow
solution with maximal supersymmetry,'' JHEP \textbf{0404}, 011 (2004)
[arXiv:hep-th/0304132]. 

\bibitem{Bena:2004jw} I.~Bena and N.~P.~Warner, ``A harmonic family of
dielectric flow solutions with maximal supersymmetry,'' JHEP \textbf{0412},
021 (2004) [arXiv:hep-th/0406145]. 

\bibitem{Lin:2004nb} H.~Lin, O.~Lunin and J.~M.~Maldacena, ``Bubbling AdS
space and 1/2 BPS geometries,'' JHEP \textbf{0410}, 025 (2004)
[arXiv:hep-th/0409174];

\bibitem{Lin:2005nh} H.~Lin and J.~M.~Maldacena, ``Fivebranes from gauge
theory,'' Phys.\ Rev.\ D \textbf{74}, 084014 (2006) [arXiv:hep-th/0509235]. 


\bibitem{Hosomichi:2008jb} K.~Hosomichi, K.~M.~Lee, S.~Lee, S.~Lee and
J.~Park, ``N=5,6 Superconformal Chern-Simons Theories and M2-branes on
Orbifolds,'' arXiv:0806.4977 [hep-th]. 

\bibitem{Hosomichi:2008jd} K.~Hosomichi, K.~M.~Lee, S.~Lee, S.~Lee and
J.~Park, \textquotedblleft N=4 Superconformal Chern-Simons Theories with
Hyper and Twisted Hyper Multiplets,\textquotedblright\ arXiv:0805.3662
[hep-th]; 

D.~Gaiotto and E.~Witten, ``Janus Configurations, Chern-Simons Couplings,
And The Theta-Angle in N=4 Super Yang-Mills Theory,'' arXiv:0804.2907
[hep-th]. 

\bibitem{Ho:2008nn} P.~M.~Ho and Y.~Matsuo, ``M5 from M2,'' arXiv:0804.3629
[hep-th]; 

P.~M.~Ho, Y.~Imamura, Y.~Matsuo and S.~Shiba, ``M5-brane in three-form flux
and multiple M2-branes,'' arXiv:0805.2898 [hep-th]; 

H.~Awata, M.~Li, D.~Minic and T.~Yoneya, ``On the quantization of Nambu
brackets,'' JHEP \textbf{0102}, 013 (2001) [arXiv:hep-th/9906248]; 

J.~H.~Park and C.~Sochichiu, \textquotedblleft Single M5 to multiple M2:
taking off the square root of Nambu-Goto action,\textquotedblright\
arXiv:0806.0335 [hep-th];

K.~Furuuchi, S.~Y.~Shih and T.~Takimi, ``M-Theory Superalgebra From Multiple
Membranes,'' arXiv:0806.4044 [hep-th]; 

I.~A.~Bandos and P.~K.~Townsend, \textquotedblleft Light-cone M5 and
multiple M2-branes,\textquotedblright\ arXiv:0806.4777 [hep-th]; 

C.~S.~Chu, P.~M.~Ho, Y.~Matsuo and S.~Shiba, ``Truncated Nambu-Poisson
Bracket and Entropy Formula for Multiple Membranes,'' arXiv:0807.0812
[hep-th]; 

N.~Kim, \textquotedblleft How to put the Bagger-Lambert theory on an
orbifold : A derivation of the ABJM model,\textquotedblright\
arXiv:0807.1349 [hep-th]. 


\bibitem{Gustavsson:2008bf} A.~Gustavsson, \textquotedblleft One-loop
corrections to Bagger-Lambert theory,\textquotedblright\ arXiv:0805.4443
[hep-th];

J.~Bedford and D.~Berman, ``A note on Quantum Aspects of Multiple
Membranes,'' arXiv:0806.4900 [hep-th]. 

\bibitem{Krishnan:2008zm} D.~S.~Berman and N.~B.~Copland, ``Five-brane
calibrations and fuzzy funnels,'' Nucl.\ Phys.\ B \textbf{723}, 117 (2005)
[arXiv:hep-th/0504044]; 

D.~S.~Berman and N.~B.~Copland, \textquotedblleft A note on the M2-M5 brane
system and fuzzy spheres,\textquotedblright\ Phys.\ Lett.\ B \textbf{639},
553 (2006) [arXiv:hep-th/0605086]; 

C.~Krishnan and C.~Maccaferri, ``Membranes on Calibrations,''
arXiv:0805.3125 [hep-th]; 

I.~Jeon, J.~Kim, N.~Kim, S.~W.~Kim and J.~H.~Park, ``Classification of the
BPS states in Bagger-Lambert Theory,'' arXiv:0805.3236 [hep-th]; 

F.~Passerini, \textquotedblleft M2-Brane Superalgebra from Bagger-Lambert
Theory,\textquotedblright\ arXiv:0806.0363 [hep-th]. 
\end{thebibliography}
\end{document}